\documentclass[journal,comsoc]{IEEEtran}
\usepackage[T1]{fontenc}% optional T1 font encoding

\usepackage{ucs}
\usepackage{cite}
\usepackage{amsmath,amssymb,amsfonts}
\usepackage{algorithmic}
\usepackage{graphicx}
\usepackage{textcomp}
\usepackage{xcolor}
\usepackage{ucs}
\usepackage[utf8x]{inputenc}
\usepackage{stackrel}
\usepackage{cite, amsfonts, amssymb, amsthm, bm, bbm, graphicx, relsize, multirow, booktabs, tikz,subfigure,soul}
\usepackage[american]{babel}
\usepackage{tabulary}
\usepackage{algorithmic, algorithm}
\usepackage[multiple]{footmisc}
\theoremstyle{definition}

\theoremstyle{remark}

\newcommand{\ds}{\displaystyle}
\newcommand{\norm}[1]{\left\lVert#1\right\rVert}

\ifCLASSINFOpdf
\else
\fi
\usepackage{amsmath}
\usepackage[cmintegrals]{newtxmath}
\begin{document}
\bstctlcite{IEEEexample:BSTcontrol}
\title{Improving Cell-Free Massive MIMO by \\ Local Per-Bit Soft  Detection}

\author{Carmen D'Andrea, \IEEEmembership{Member, IEEE} and Erik G. Larsson, \IEEEmembership{Fellow, IEEE}
	\thanks{C. D'Andrea is with the Department of Electric and Information Engineering (DIEI), University of Cassino and Southern Latium, 03043 Cassino, Italy, and with Consorzio Nazionale Interuniversitario per le Telecomunicazioni (CNIT), 43124 Parma, Italy, (e-mail: carmen.dandrea@unicas.it) and E. G. Larsson is with the Department of Electrical Engineering (ISY), Link\"{o}ping University, 58183 Link\"{o}ping, Sweden, (e-mail: erik.g.larsson@liu.se). This work was developed during the visiting period of the first author to Link\"{o}ping University. The work of C. D'Andrea has been supported by the MIUR Project ``Dipartimenti di Eccellenza 2018-2022'', the MIUR PRIN 2017 Project ``LiquidEdge'' and the ``Starting Grant 2020'' (PRASG) Research Project. The work of E. G. Larsson was partially supported by VR and ELLIIT. }}


\markboth{IEEE Communications Letters,~Vol.-, No.-, April~2021}%
{D'Andrea  \MakeLowercase{\textit{and}} Larsson: Improving Cell-Free Massive MIMO by Local Per-Bit Soft  Detection}



\maketitle

\begin{abstract}
In this letter, we consider the uplink of a cell-free Massive multiple-input multiple-output (MIMO) network where each user is decoded by a subset of access points (APs). An additional step is introduced in the cell-free Massive MIMO processing: each AP in the uplink locally implements soft MIMO detection and then shares the resulting bit log-likelihoods on the front-haul link. The decoding of the data is performed at the central processing unit (CPU), collecting the data from the APs. 
The non-linear processing at the APs consists of the approximate computation of the posterior density for each received data bit, exploiting only  local channel state information. The proposed method offers good performance in terms of frame-error-rate and considerably lower complexity than the optimal maximum-likelihood demodulator.
\end{abstract}


% \fi
\IEEEpeerreviewmaketitle

\section{Introduction}
\IEEEPARstart{C}{ell-free} Massive MIMO systems consist of a very large number of distributed APs serving many users in the same time-frequency resource \cite{Ngo_CellFree2017}. In a cell-free Massive MIMO system the APs locally estimate the channels towards all the users and then use these estimates to transmit/decode data using a linear processing. 
All the APs are connected to a CPU and cooperate via a front-haul network, serving the users in time-division-duplex (TDD) operation, so that  there are actually no cell boundaries.
The cell-free Massive MIMO concept is a recent research topic that has been gaining huge attention in the last few years. 
The assumption that all the APs serve all the users in the system makes the system unscalable and it is pointless to waste power and computational resources at an AP to decode users that are very far away and that are received with a very low signal-to-interference-noise-ratio (SINR). 
A user-centric (UC) approach to the cell-free Massive MIMO is considered in \cite{Buzzi_DAndrea_Zappone_TWC2019}, where each user is served only by the APs that are in its immediate vicinity. 

The first papers on cell-free Massive MIMO considered maximum-ratio transmission/detection implemented locally at the APs. 
Recently other literature proposed to improve the performance by using more sophisticated precoding and combining schemes implemented locally at the APs in order to facilitate a scalable implementation. 
Specifically, in \cite{Attarifar_WCL2019} the authors propose a modification of conjugate beamforming for the downlink which eliminates the self-interference, and in  \cite{Buzzi_DAndrea_ISWCS2018} and \cite{interdonato2019localZF} partial zero-forcing on the downlink and  successive interference cancellation on the uplink are considered. The results reveal a significant performance improvement over simple maximum ratio transmission/detection. Moreover, \cite{Zakir_Larsson_RadioStripes_2020} considered sequential processing algorithm with normalized linear
minimum mean square error combining at every AP assuming the radio stripe network architecture.
Additionally,  \cite{Bashar_NOMAOMA_TCOM2019} introduced a non-orthogonal-multiple-access (NOMA)-based cell-free Massive MIMO system with successive interference cancellation implemented at the users' sides. A fully decentralized architecture for co-located Massive MIMO uplink based on recursive methods is presented in \cite{Edfors_SiPS2018}, where the authors propose algorithms providing a sequence of estimates that converge asymptotically to the zero-forcing solution.

The main insight behind this letter is that the linear per-AP processing used in previous work on decentralized detection \cite{Ngo_CellFree2017,Buzzi_DAndrea_Zappone_TWC2019,Attarifar_WCL2019,Buzzi_DAndrea_ISWCS2018,interdonato2019localZF,Zakir_Larsson_RadioStripes_2020} is highly suboptimal, and could even be ill-conditioned. We propose to improve the performance of cell-free Massive MIMO by employing an intermediate \emph{non-linear step} based on  \emph{locally-implemented} soft MIMO detection. Soft MIMO detection for point-to-point MIMO systems is a well investigated topic in the literature, and many algorithms exist with different performance-complexity
tradeoffs \cite{Larsson_PartialMarginalization2008,Choi_TSP2010}. The optimal soft detector for point-to-point MIMO is also well known \cite{Hochwald_NearCapacity2003}.

It should be noted that cell-free Massive MIMO, cloud radio access network (C-RAN) and coordinated multi-point transmission are all instances of a distributed MIMO architecture \cite{CRAN_CST2014,CoMP_CST2017}. In such systems, one may either collect all baseband data to a central unit for processing, or one may distribute some of the processing at the different access points. The motivation of distributing some of the processing is to reduce implementation complexity and specifically reduce on the required fronthaul signaling. In the literature, various distributed processing schemes are proposed and compared. However, to the best of our knowledge, only \emph{linear} schemes have been investigated. In this context, our contribution is to propose and initially investigate the use of \emph{non-linear} processing per access point before the corresponding data are forwarded to a CPU.

\subsection*{Contribution}
\noindent
In this paper, we focus on the uplink of a cell-free Massive MIMO system considering local non-linear processing at each AP before sharing the local estimates on the front-haul link. Specifically, an additional step is introduced in the processing where each AP in the uplink performs \textit{local soft MIMO detection}. Each AP in the uplink locally (a) estimates the channels towards the users, (b) collects the received signals on the uplink (c) performs the local soft MIMO detection of the data and, (d) sends the resulting bit log-likelihoods on the front-haul link to the CPU. For each user, the log-likelihood ratios (LLR) computed at the APs are shared on the front-haul link and then the CPU  collects LLRs from the APs and decodes the data. We assume that each user is served by a subset of APs in the network and the LLRs are computed using the fixed-complexity partial marginalization (PM) detector of  \cite{Larsson_PartialMarginalization2008}. This algorithm offers an attractive performance-complexity trade-off and is suitable for highly parallel hardware. Numerical results reveal that the local detection based on the PM outperforms the maximum-ratio combining (MRC), the zero-forcing with decision feed (ZF-DF) and the minimum-mean-square-error successive interference cancellation (MMSE-SIC) and gives performance comparable to the exact maximum-likelihood (ML) detector but with a considerably lower complexity.

\section{System model and channel estimation}
A network that consists of $M$ APs, equipped with a uniform linear array (ULA) with $N_{\rm AP}$ antennas, and $K$ single-antenna users is here considered. The $M$ APs are connected by means of a front-haul network to a CPU wherein data-decoding is performed. 
We denote by $\mathbf{g}_{k,m} \sim \mathcal{CN}\left( \mathbf{0}_{N_{\rm AP}}, \mathbf{R}_{k,m} \right)$ the channel between the $k$-th user and the $m$-th AP, where $\mathbf{R}_{k,m} \in \mathbb{C}^{N_{\rm AP} \times N_{\rm AP}}$ is the spatial correlation matrix, which describes the spatial properties of the channel and $\beta_{k,m}=\text{tr}\left( \mathbf{R}_{k,m}\right)/N_{\rm AP} $ is the large-scale fading coefficient that describes geometric path-loss and shadowing.
The dimension in time/frequency samples of the channel coherence length is denoted by $\tau_c$, and the dimension of the uplink training phase by $\tau_p < \tau_c$. The pilot sequences transmitted by the users, $\bm{\phi}_k, \, k=1,\ldots, K$, are chosen in the set of $\tau_p$ orthonormal sequences, if $\tau_p<K$ pilot contamination degrades the system performance. The $m$-th AP estimates the channel vector $\mathbf{g}_{k,m}$ based on the observable given by the projection of the received signal on the pilot sequence assigned to the $k$-th user, i.e., 
\begin{equation}
\widehat{\mathbf{y}}_{k,m}=\sqrt{p_k}\mathbf{g}_{k,m} + \ds \sum_{\substack{i=1 \\ i\neq k}}^K {\sqrt{p_i}\mathbf{g}_{i,m}\boldsymbol{\phi}_i^H \boldsymbol{\phi}_k} + \mathbf{w}_{k,m} \; ,
\label{y_hat_ka}
\end{equation}
where ${p}_k=\tau_p \widetilde{p}_k$ denotes the power employed by the $k$-th user during the training phase, $\widetilde{p}_k$ is the power transmitted for each sample of the pilot sequence used by the $k$-th user, and $\mathbf{w}_{k,m} $ contains the thermal noise contribution with i.i.d. entries ${\cal CN}(0, \sigma^2_w)$. The MMSE channel estimate of the channel $\mathbf{g}_{k,m}$ can be written as 
$
\hat{\mathbf{g}}_{k,m}= \sqrt{p_k}\mathbf{R}_{k,m} \bm{\Gamma}_{k,m}^{-1}\widehat{\mathbf{y}}_{k,m} \;,
$
where $\bm{\Gamma}_{k,m}= \sum_{i=1}^K p_i\mathbf{R}_{i,m}|\boldsymbol{\phi}_i^H\boldsymbol{\phi}_k|^2 + \sigma^2_w\mathbf{I}_{N_{\rm AP}}$.
The channel estimate is distributed as $\hat{\mathbf{g}}_{k,m} \sim \mathcal{CN}\left(\mathbf{0}_{N_{\rm AP}}, p_k\mathbf{R}_{k,m} \bm{\Gamma}_{k,m}^{-1}\mathbf{R}_{k,m}\right)$, and the channel estimation error is $\widetilde{\mathbf{g}}_{k,m}=\mathbf{g}_{k,m} - \hat{\mathbf{g}}_{k,m}$ and distributed as $\widetilde{\mathbf{g}}_{k,m} \sim \mathcal{CN}\left(\mathbf{0}_{N_{\rm AP}}, \mathbf{C}_{k,m} \right)$, with $\mathbf{C}_{k,m}= \mathbf{R}_{k,m}- p_k\mathbf{R}_{k,m} \bm{\Gamma}_{k,m}^{-1}\mathbf{R}_{k,m}$.

In the uplink data decoding we assume an AP-centric approach, i.e., the $m$-th AP serves the $N_m$ users that it receives with best average channel conditions. Let $S_m \, : \, \{1,\ldots, K \} \rightarrow \{1,\ldots, K \}$ denote the sorting operator for the vector $\left[\beta_{1,m},\ldots, \beta_{K,m}\right]$, such that $\beta_{S_m(1),m} \geq \beta_{S_m(2),m} \geq \ldots \geq \beta_{S_m(K),m}$. The set $\mathcal{K}_m$ of the $N_m$ MSs served by the $m$-th AP is then given by
$
\mathcal{K}_m=\{ S_m(1), S_m(2), \ldots , S_m(N_m) \}.
$
Consequently, the set of APs serving the $k$-th user is defined as  $\mathcal{M}_k=\{ m: \,  k \in \mathcal{K}_m \}$\footnote{Several user-association schemes can be considered, for example, a UC approach in \cite{Buzzi_DAndrea_Zappone_TWC2019}, or performance-maximizing association rules, by defining the sets $\mathcal{K}_m$ and $\mathcal{M}_k$ accordingly.}.

\section{Uplink non-linear processing} \label{NL_processing}
In uplink, users send their data symbols without any channel-dependent phase offset. As a result, the signal received at the $m$-th AP in the generic symbol interval can be expressed as 
\begin{equation}
{\overline{\mathbf{y}}}_m=\ds \sum_{k =1}^K \ds \sqrt{\eta_{k}} \mathbf{g}_{k,m} x_k+ \overline{\mathbf{w}}_m \; ,
\label{y_m_received}
\end{equation}
with ${\eta_{k}}$ and $x_k$ representing the uplink transmit power and the data symbol of the $k$-th user in the generic symbol interval, respectively, and $\overline{\mathbf{w}}_m \sim {\cal CN}\left(\mathbf{0}_{N_{\rm AP}}, \sigma^2_w \mathbf{I}_{N_{\rm AP}} \right)$ the noise vector.

We define the following vectors and matrices $$\overline{\mathbf{x}}_{\mathcal{S}}=\left[  x_{\mathcal{S}(1)}, \ldots, x_{\mathcal{S}(S)} \right]^T,$$ for a generic set $\mathcal{S}$ with cardinality $S$, 
$$\widehat{\overline{\mathbf{B}}}_{\mathcal{K}_m,m}= \left[ \sqrt{\eta_{\mathcal{K}_m(1)}} \widehat{\mathbf{g}}_{\mathcal{K}_m(1),m}, \ldots,  \sqrt{\eta_{\mathcal{K}_m(N_m)}} \widehat{\mathbf{g}}_{\mathcal{K}_m(N_m),m} \right],$$ contains the power control coefficients and the channel estimates for the users in $\mathcal{K}_m$, $$\widetilde{\overline{\mathbf{B}}}_{\mathcal{K}_m}= \left[ \sqrt{\eta_{\mathcal{K}_m(1)}} \widetilde{\mathbf{g}}_{\mathcal{K}_m(1),m}, \ldots,  \sqrt{\eta_{\mathcal{K}_m(N_m)}} \widetilde{\mathbf{g}}_{\mathcal{K}_m(N_m),m} \right],$$ contains the power control coefficients and the channel estimation errors for the users in $\mathcal{K}_m$, and similarly $\overline{\mathbf{B}}_{\overline{\mathcal{K}}_m}$ contains the power control coefficients and the true channels for the users in the set  $ \overline{\mathcal{K}}_m = \lbrace k \notin \mathcal{K}_m \rbrace$.
Given the previous definitions, we rewrite $\mathbf{\overline{y}}_m$ as follows:
\begin{equation}
\begin{array}{llll}
\mathbf{\overline{y}}_m=&\widehat{\overline{\mathbf{B}}}_{\mathcal{K}_m} \overline{\mathbf{x}}_{\mathcal{K}_m} + \widetilde{\overline{\mathbf{B}}}_{\mathcal{K}_m} \overline{\mathbf{x}}_{\mathcal{K}_m} + \overline{\mathbf{B}}_{\overline{\mathcal{K}}_m} \overline{\mathbf{x}}_{\overline{\mathcal{K}}_m} + \overline{\mathbf{w}}_m \, .
\end{array}
\label{y_m_received2}
\end{equation}
Note that in Eq. \eqref{y_m_received2} the $m$-th AP knows the channel estimates for the users in $\mathcal{K}_m$, i.e., the matrix $\widehat{\overline{\mathbf{B}}}_{\mathcal{K}_m}$, and performs the non-linear processing only for these users.

We define the interference-plus-noise contribution at the $m$-th AP as 
\begin{equation}
\overline{\mathbf{e}}_m= \widetilde{\overline{\mathbf{B}}}_{\mathcal{K}_m} \overline{\mathbf{x}}_{\mathcal{K}_m} + \overline{\mathbf{B}}_{\overline{\mathcal{K}}_m} \overline{\mathbf{x}}_{\overline{\mathcal{K}}_m} + \overline{\mathbf{w}}_m \, ,
\end{equation}
where the $m$-th AP treats the channel estimates of the users in $\mathcal{K}_m$ as true channel and considers as interference both the channel estimation errors for users in $\mathcal{K}_m$ and the channels of the users in in $\overline{\mathcal{K}}_m$. We use the Gaussian approximation of the interference plus noise terms in $\overline{\mathbf{e}}_m$, i.e., we assume $\overline{\mathbf{e}}_m \sim \mathcal{CN} \left( \mathbf{0}_{N_{\rm AP}}, \sigma^2_{e,m} \mathbf{I}_{N_{\rm AP}}\right)$, with
\begin{equation}
\begin{array}{llll}
\sigma^2_{e,m}=& \ds \sum_{j \in \mathcal{K}_m} {\eta_j \frac{\text{tr}\left(\mathbf{C}_{j,m}\right)}{N_{\rm AP}} }  + \ds \sum_{j \in \overline{\mathcal{K}}_m} {\eta_j\frac{\text{tr}\left(\mathbf{R}_{j,m}\right)}{N_{\rm AP}} } + \sigma^2_w \, .
\end{array}
\end{equation} 

Eq. \eqref{y_m_received2} can be thus written as
$
\mathbf{\overline{y}}_m\!=\!\widehat{\overline{\mathbf{B}}}_{\mathcal{K}_m} \overline{\mathbf{x}}_{\mathcal{K}_m} \!+\! \overline{\mathbf{e}}_m \, .
$

We focus on a real-valued discrete-time matrix-vector model and assume that in $\mathbf{\overline{y}}_m$ the real and imaginary part represent the inphase and quadrature components of the signal, respectively, i.e.,
\begin{equation}
\begin{array}{llll}
\mathbf{{y}}_m=\widehat{\mathbf{B}}_{\mathcal{K}_m} \mathbf{x}_{\mathcal{K}_m} + \mathbf{e}_m \, .
\end{array}
\label{y_m_real}
\end{equation}
The interference-plus-noise vector $\mathbf{e}_m$ has independent Gaussian elements with zero mean and variance $\sigma^2_{e,m}/2$, hence, we can write the distribution of the received signal $\mathbf{y}_m$ given the matrix $\widehat{\mathbf{B}}_{\mathcal{K}_m}$ and the symbols $\mathbf{x}_{\mathcal{K}_m}$ as 
\begin{equation}
\begin{array}{llll}
p\left( \mathbf{{y}}_m | \widehat{\mathbf{B}}_{\mathcal{K}_m},\mathbf{x}_{\mathcal{K}_m} \right) =& \ds \frac{1}{\sqrt{\pi^{2N_m} \sigma_{e,m}^{4 N_m}}} \\ & \times \text{exp} \left(  -\ds \frac{1}{\sigma^2_{e,m}} \norm{\mathbf{{y}}_m- \widehat{\mathbf{B}}_{\mathcal{K}_m}\mathbf{x}_{\mathcal{K}_m}  }^2 \right) 
\end{array}
\label{pdf_y}
\end{equation}
The vector $\mathbf{x}_{\mathcal{K}_m}$ in Eq. \eqref{y_m_real} has elements that belong to a finite alphabet $\mathcal{A}$ and each entry, say $s_{\ell}$, is composed of $q$ information bits, hence, the vector $\mathbf{x}_{\mathcal{K}_m}$ is composed of $2 N_m q$ bits, $b_1, \ldots, b_{2 N_m q}$ say, assumed independent. To each bit $b_i$ we associate an \textit{a-priori} LLR is 
\begin{equation}
L(b_i)=\log \left( \ds \frac{P(b_i=1)}{P(b_i=0)}\right) \, ,
\end{equation}
which expresses what the detector at the generic AP knows about the bit before the data $\mathbf{y}_m$ are observed.

\begin{figure*}
	\begin{equation}
	L(b_i | \mathbf{y}_m,\widehat{\mathbf{B}}_{\mathcal{K}_m} ) = \log \left( \ds \ds \sum_{\mathbf{x}_{\mathcal{K}_m} : b_i(\mathbf{x}_{\mathcal{K}_m})=1}  \text{exp} \left(  -\ds \frac{1}{\sigma^2_{e,m}} \norm{\mathbf{{y}}_m- \widehat{\mathbf{B}}_{\mathcal{K}_m}\mathbf{x}_{\mathcal{K}_m}  }^2 \right) \right) - \log \left( \sum_{\mathbf{x}_{\mathcal{K}_m} : b_i(\mathbf{x}_{\mathcal{K}_m})=0 } \text{exp} \left(  -\ds \frac{1}{\sigma^2_{e,m}} \norm{\mathbf{{y}}_m- \widehat{\mathbf{B}}_{\mathcal{K}_m}\mathbf{x}_{\mathcal{K}_m}  }^2 \right)\right)
	\label{LLR_given_y}
	\end{equation}
	\hrulefill
\end{figure*}

Assuming that all the bits are equally likely to be 0 or 1 before observing $\mathbf{y}_m$, the LLR given the observable $\mathbf{y}_m$ is written as in Eq. \eqref{LLR_given_y} at the top on the next page\cite{Larsson_PartialMarginalization2008}.

\subsection{Detection via partial marginalization}
We consider the method of PM to compute \eqref{LLR_given_y}. In the following we report the basic idea of the PM method and refer to paper \cite{Larsson_PartialMarginalization2008} for further details on the procedure. The idea behind the PM is to perform a two-step marginalization of the posterior density for $\mathbf{y}_m$, performing exact marginalization over a carefully chosen, fixed number, say $r_m$, of the $2N_m q$ bits and to approximately marginalize over the remaining $2N_mq-r_m$ bits, using the max-log philosophy. 
Let $\mathcal{B}$ be a bit index permutation on $[1, \ldots , 2N_mq]$, we marginalize Eq. \eqref{LLR_given_y} \textit{exactly} over $b_{\mathcal{B}_1}, \ldots, b_{\mathcal{B}_{r_m}}$ and \textit{approximately}, using the max-log philosophy, over $b_{\mathcal{B}_{r_m+1}}, \ldots, b_{\mathcal{B}_{2N_mq}}$. The max-log approach approximates each of the sums with their largest term but it is still subject to the constellation constraint on $\mathbf{x}_{\mathcal{K}_m}$ and requires the solution of NP-hard maximization problems. To overcome this issue the PM method uses computationally less expensive approximations provided by the hard ZF-DF detector. The PM approach is thus composed of two approximations: (i) replacing the marginalization over $b_{\mathcal{B}_{r_m+1}}, \ldots, b_{\mathcal{B}_{2N_mq}}$ by a max-log operation and (ii) solving this max-log problem approximately using a low-complexity method based on ZF-DF detector.

\subsection{Sharing the LLRs on the front-haul link} \label{Sharing_LLRs}
The computation of Eq. \eqref{LLR_given_y} via PM is locally implemented at each AP, $\forall \, i=1,\ldots,2N_mq $ and $\forall \, m=1,\ldots,M$. These values are shared on the front-haul link and the CPU decodes the generic bit transmitted by the $k$-th user collecting the LLRs provided from the APs decoding the $k$-th user. Otherwise stated, assume that the $k$-th users transmits $n$ information bits $b_1^{(k)}, \ldots b_n^{(k)}$, the decoding on the bit $b_i^{(k)}$ is obtained by\footnote{According to the well-known Bayes rule and assuming independent observations at the APs, considering the sum of LLRs is optimal.}
\begin{equation}
f\left( b_i^{(k)} \right) = \ds \sum_{m \in \mathcal{M}_k} L(b_i^{(k)} | \mathbf{y}_m,\widehat{\mathbf{B}}_{\mathcal{K}_m} ) \stackrel[\widehat{b}_i^{(k)}=0]{\widehat{b}_i^{(k)}=1}{\gtrless} 0 .
\end{equation}

\subsection{Complexity}
Following \cite{Larsson_PartialMarginalization2008}, the number of operations per bit of the PM at the $m$-th AP is $O(4 N_m^2 2^{r_m})$, while in exact demodulation \eqref{LLR_given_y} it is $O(2^{2N_mq})$. In the proposed approach, this procedure should be performed at each AP for the data transmitted by the users in $\mathcal{K}_m$. For the $k$-th user, the APs in $\mathcal{M}_k$ perform the non-linear processing based on the PM procedure and share on the front-haul link the LLRs computed for the $k$-th user's bits. The CPU receives all the LLRs from the APs in $\mathcal{M}_k$ and decodes the data transmitted by the $k$-th user. Thus, we can observe that for the $k$-th user, the number of operations per bit in our approach is $O(|\mathcal{M}_k|4 N_m^2 2^{r_m})$, while in the exact demodulation it is $O( |\mathcal{M}_k|2^{2N_mq})$, neglecting the sum of the LLRs at the CPU in both the cases. 

While the computational complexity of the proposed approach is higher compared to linear local processing, as we will see in the numerical results, the gain in performance is significant. As an alternative to the specific non-linear processing (partial marginalization) proposed here, using deep-learning methods instead  might have potential and is something that could be investigated in the future.  For example, the techniques in \cite{Hoydis_DL_TGCN2017} might be applicable. In this context it should also be stressed that eventual decoding performance is not the only important aspect. For example, the partial marginalization algorithm is known to have fixed complexity, hence enabling the design of ultra-efficient FPGA's or application specific integrated circuits. In this respect, there could be value in using non-linear processing per AP that actually relies on well-established technology and hardware implementations. For example, one implementation of SUMIS, a variation on the partial marginalization theme \cite{SUMIS_Algorithm2014}, was developed in \cite{Frostensson2013}.

\section{Numerical Results}
We consider a square area of 1 km$^2$ wrapped around at the edges to avoid boundary effects. We assume $M=50$ APs each with an 8-element ULA with $\lambda/2$ spacing, i.e., $N_{\rm AP} = 8$, and single antenna users. The communication bandwidth is $W = 20$ MHz centered over the carrier frequency $f_0=1.9$ GHz, the power spectral density (PSD) of the noise is -174 dBm/Hz and the noise figure at the receiver is 9 dB. With regard to the channels from users to the APs, we assume $\mathbf{R}_{k,m} = \beta_{k,m} \mathbf{I}_{N_{\rm AP}}$\footnote{The case of correlated channels was also considered obtaining the same qualitative relation between the performance.}. The LSF coefficient $\beta_{k,m}$ in dB is modelled as in \cite[Table B.1.2.2.1-1]{3GPP_36814_GUE_model}. The shadow fading coefficients from an AP to different users are correlated and follows \cite[Table B.1.2.2.1-4]{3GPP_36814_GUE_model}. We assume knowledge of the LSF coefficients at the CPU and the association between users and APs is performed at the CPU. The length of the channel estimation phase is $\tau_p=12$ samples and each user transmits 100 mW during the uplink training, i.e., $\widetilde{p}_k=100$ mW, $\forall \; k=1,\ldots, K$.
Fractional power control (FPC) is assumed during the uplink data transmission, the transmit power of the $k$-th user is
$
\eta_k^{\rm UL}= \text{min} \left( P_{{\rm max},k}, P_0 \zeta_k^{-\kappa}\right)\, , 
$
where $P_{{\rm max},k}$ is the maximum $k$-th user transmit power, $P_0$ is a specific parameter configurable by the serving APs, $\kappa$ is a path loss compensation factor, and
$$
\zeta_k= \sqrt{\ds \sum_{m \in \mathcal{M}_k} {\beta_{k,m}}} \, .
$$
In the simulations, we use $P_{{\rm max},k}=100$mW $\forall \; k$, $P_0=-10$dBmW and $\kappa=0.5$.
We present numerical results in terms of frame-error-rate (FER) to illustrate the performance of the proposed approach. Monte Carlo simulation was used to simulate the FER and at each signal-to-noise-ratio (SNR) point, we simulate enough frame to count 200 frame errors. We present the FER performance of the $k$-th user, positioned at the centre of the simulation area, as a function of $\text{SNR}_k$ defined as
\begin{equation}
\text{SNR}_k=\frac{\eta_k N_{\rm AP} \ds \sum_{m \in \mathcal{M}_k}{\beta_{k,m}}}{\sigma^2_w}\, .
\end{equation}
We assume QPSK modulation, i.e., $q=1$ in Section \ref{NL_processing}. In the following results, we assume that all the APs serve the same number of users, i.e., $N_m=N, \forall \, m=1,\ldots, M$ and that the parameter of the PM is the same for all the APs, i.e., $r_m=r, \forall \, m=1,\ldots,M$.

\begin{figure}[!t]
	\centering
	\includegraphics[scale=0.38]{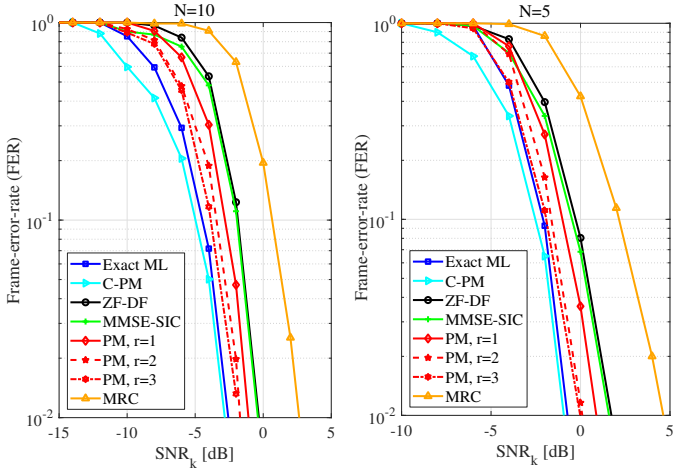}
	\caption{Performance comparison in terms of FER. Parameters: $M=50$, $K=20$, $N_{\rm AP}=8$ $\tau_p=12$.}
	\label{Fig:FER_K20}
\end{figure}

\begin{figure}[!t]
	\centering
	\includegraphics[scale=0.38]{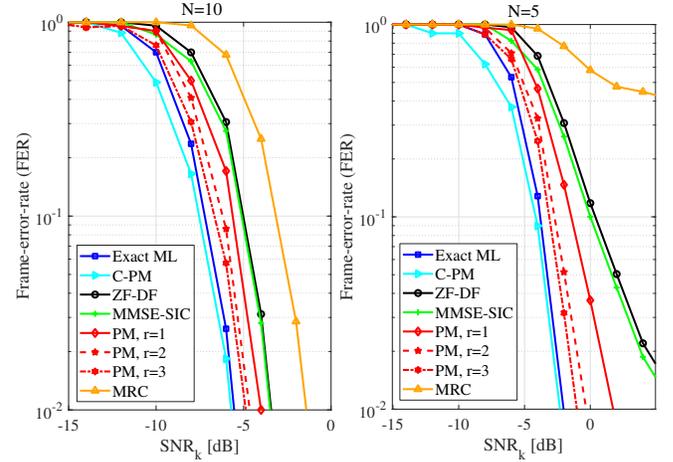}
	\caption{Performance comparison in terms of FER. Parameters: $M=50$, $K=30$, $N_{\rm AP}=8$ $\tau_p=12$.}
	\label{Fig:FER_K30}
\end{figure}

In Figs. \ref{Fig:FER_K20} and \ref{Fig:FER_K30}, we report the performance in terms of FER of local PM with different values of $r$, ZF-DF with V-BLAST ordering, MMSE-SIC, centralized implementation of PM at the CPU (C-PM), simple MRC implemented at the APs, and exact ML in \eqref{LLR_given_y} with $K=20$ and $K=30$, respectively. Coded transmission is assumed and each codeword spans one realization of the channels. We used a convolutional code with block length 100 bits and a rate 1/3 as outer code decoded with the Viterbi algorithm with no iteration between the decoder and the demodulator. First of all, we can see that increasing the parameter $N$, i.e., the number of users served by each AP, in both the figures, the performance in terms of FER considerably improves. We can also see that the presence of a larger number of users in the system, decreases the performance, especially for the MRC. This is due to the increase of the variance of the interference-plus-noise contribution. Higher numbers of the parameter $r$ make the performance of the PM closer to the exact demodulation, with a gain in computational complexity. We can also see that the local PM offers better performance with respect to the ZF-DF and MMSE-SIC. The C-PM offers the better performance compared with the local alternatives because it requires a \textit{complete sharing} of the channel estimates on the front-haul link.

\section{Conclusions}
In this work, we considered the uplink of a cell-free Massive MIMO system with non-linear processing at each AP. 
An additional step is introduced at the APs which locally implements a non-linear soft MIMO detector  then shares the so-obtained per-bit log-likelihood ratios on the front-haul link. 
We assume that each AP decodes a subset of users in the system. The decoding of the data  is performed at the CPU, by collecting soft bits  from the APs for each user. 
The soft MIMO detector at the APs is based on the PM algorithm \cite{Larsson_PartialMarginalization2008}, and computes  the posterior density for the received data bits exploiting only  local  channel state information. 
Numerical results show the effectiveness of the proposed approach, that gives performance close to that of  exact demodulation with a significantly lower complexity, offering a considerable improvement with respect to the traditional approaches in cell-free Massive MIMO.

\bibliography{Cell_free_references}
\bibliographystyle{ieeetran}

\end{document}